\documentclass[a4paper,fleqn,usenatbib]{mnras}

\usepackage[T1]{fontenc}
\usepackage{ae,aecompl}

\usepackage{graphicx}	
\usepackage{amsmath}	
\usepackage{amssymb}	

\def\arcmin{\hbox{$^\prime$}}

\def\lum{erg s$^{-1}$}

\title[Spin evolution in LMC\,X-4]{Near-periodical spin period evolution in the binary system LMC\,X-4}

\author[S. Molkov et al.]{
  S. Molkov$^{1}$\thanks{E-mail: molkov@iki.rssi.ru},
  A. Lutovinov$^{1,2}$,
  M. Falanga$^{3}$,
  S. Tsygankov$^{4}$
  and E. Bozzo$^{5}$
  \\
  $^{1}$ Space Research Institute, Russian Academy of Sciences, Profsoyuznaya 84/32, 117997 Moscow, Russia\\
  $^2$ Moscow Institute of Physics and Technology, Institutskiy per. 9, Dolgoprudny,
Moscow Region, 141700, Russia \\
  $^{3}$ International Space Science Institute (ISSI), Hallerstrasse 6, CH-3012 Bern, Switzerland\\
  $^{4}$ Tuorla Observatory, Department of Physics and Astronomy,
  University of Turku,  V\"ais\"al\"antie 20, FI-21500, Piikki\"o, Finland\\
  $^{5}$ ISDC, University of Geneva, chemin d'Ecogia 16, 1290, Versoix, Switzerland}

\date{Accepted 2016 September 22. Received 2016 September 21; in original
form 2016 July 7}

\pubyear{2016}

\begin{document}
\label{firstpage}
\pagerange{\pageref{firstpage}--\pageref{lastpage}}
\maketitle

\begin{abstract}
In this paper we investigated the long-term evolution of the pulse-period in
the high-mass X-ray binary LMC\,X-4 by taking advantage of more than 43~yrs
of measurements in the X-ray domain. Our analysis revealed for the first time
that the source is displaying near-periodical variations of its spin period
on a time scale of roughly 6.8~yrs, making LMC\,X-4 one of the known binary
systems showing remarkable long term spin torque reversals. We discuss
different scenarios to interpret the origin of these torque reversals.
\end{abstract}

\begin{keywords}
accretion, accretion discs -- stars: individual: LMC\,X-4 -- X-rays: binaries.
\end{keywords}



\section{Introduction}

The High-Mass X-ray Binary (HMXB) LMC\,X-4 has been observed in the X-ray
domain, as well as in other energy domains, since more than forty three
years. It was originally discovered by the {\it Uhuru} satellite
\citep[][]{giacconi72} and it is known to be located in the Large Magellanic
Cloud (LMC) at a distance of $\sim 50$~kpc. Based on the analysis of the very
deep objective-prism plates \citet{sanduleak76} proposed that the optical
counterpart of LMC\,X-4 is an OB star, showing optical flux modulations on
the time scale of $\sim1.4$~days presumably connected with the orbital motion
in the binary system \citep[][]{chevalier77, hutchings78}. Regular X-ray
eclipses occurring on the same periodicity were detected later in the X-ray
lightcurve of the source with the instruments on-board the {\it SAS-3} and
{\it Ariel\,V} observatories \citep[][]{li78,white78}, finally confirming the
binary nature of the source. The orbital parameters of the system were
investigated in a number of papers \citep[see, e.g.,][and references
therein]{levine91,safiharb96,woo96,levine00,naik04} and refined more recently
by \citet{falanga15} and \citet{molkov15}.

LMC\,X-4 is also known to display a super-orbital modulation in the X-ray
domain with a period of $P_{sup} \sim 30.5$ days, reported for the first time
by \citet{lang81} using data from the HEAO1 observatory. The dynamic range in
the X-ray luminosity displayed by the source over the period of the
super-orbital modulation (oscillating between high and low emission states)
is of about two orders of magnitude, reaching up to
$(4-5)\times10^{38}$\,\lum\ \citep[see,
e.g.][]{woo96,heemskerk89,tsygankov05,2013MNRAS.428...50G}.

One of the most widely accepted mechanisms to produce such modulation
foresees the presence of a precessing warped accretion disk around the
compact object in the binary system, which periodically obscures the source
of the X-ray radiation \citep[see, e.g.,] [and references
therein]{ogilvie01,clarkson03,kotze12}. Such scenario has also been applied
to the case of LMC\,X-4 in many papers which investigated the long term
evolution of the X-ray emission of the system \citep[see,
e.g.,][]{paul02,clarkson03,tsygankov05,molkov15}. The same super-orbital
modulation was also long known to be present in the optical waveband
\citep[][]{ilovaisky84,heemskerk89}.

LMC\,X-4 shows also sporadic large X-ray flares which have a typical duration
from tens of minutes up to few hours and occur unpredictably once every few
days. During these events the X-ray luminosity of the source can increase up
to an order of magnitude
\citep[][]{epstein77,pietsch85,levine00,moon03,tsygankov05}. The physical
mechanism triggering these flares is still poorly understood, but it is
probably related to temporarily variations in the mass accretion rate onto
the compact object \citep[][]{levine00}.

\begin{table*}
\small
\begin{center}
  \caption{Log of all LMC\,X-4 pulse-period measurements used in the present work.}
  \label{tab:history_tbl}
  \begin{tabular}{llll} 
    \hline
    \hline
    \hspace{5mm}Date & \hspace{3mm}Period & Observatory & \hspace{3mm}Reference\\
    \hspace{5mm}MJD & \hspace{6mm}(s) & $$ & $$\\
    \hline
    $42831.43^a$ & $13.528(2)$ & {\it SAS-3} & \citet{kelley83} \\
    $43286.07^a$ & $13.525(7)$ & {\it SAS-3} & \citet{kelley83} \\
    $44589.60^a$ & $13.5113(40)^b$ &{\it Einstein} & \citet{naranan85} \\
    $45656.0^c$  & $13.5019(2)$ & {\it EXOSAT} & \citet{pietsch85} \\
    $47229.3313^d$  & $13.495978(9)$ & {\it Ginga} & \citet{woo96} \\
    $47741.9904^d$  & $13.49798(1)$ & {\it Ginga} & \citet{levine91} \\
    $48558.8598^d$  & $13.50292(2)$ & {\it ROSAT} & \citet{woo96} \\
    $49468.6859^d$  & $13.5075(2)$ & {\it ASCA} & \citet{paul02a} \\
    $50227.8069^d$  & $13.5088(1)$ & {\it ASCA} & \citet{paul02a} \\
    $50315.121^d$  & $13.5085(4)$ & {\it RXTE} & this work \\
    $50743.293^d$  & $13.5068(6)$ & {\it RXTE} & this work \\
    $51106.6399^d$  & $13.50260(12)$ & {\it BeppoSAX} & \citet{naik04} \\
    $51109.457^d$  & $13.5031(3)$ & {\it RXTE} & this work \\
    $51531.984^d$  & $13.4960(4)$ & {\it RXTE} & this work \\
    $52892.1^c$  & $13.4959(5)$ & {\it XMM-Newton} & this work \\
    $53172.8^c$  & $13.4974(5)$ & {\it XMM-Newton} & this work \\
    $54480.5^c$    & $13.5087(2)$ & {\it Suzaku}  & \citet{hung10} \\
    $54507.3^c$     & $13.5091(1)$ & {\it Suzaku}  & \citet{hung10} \\
    $54561.7^c$    & $13.5096(1)$ & {\it Suzaku}  & \citet{hung10} \\
    $56112.8^c$     & $13.49892(3)$ & {\it NuSTAR} & \citet{shtykovsky16}\\
    $56996.58^a$ & $13.490(8)$ & XRT/{\it Swift} & this work\\
    \hline
  \end{tabular}

 \begin{tabular}{ll}
  $^a$  Pulsations have been detected only during flaring activity \\
  $^b$ The value has been corrected in this work, see text for details \\
  $^c$ Approximate middle time of the observation ${\hspace{0.6cm}}~$\\
  $^d$ Epoch of the mid-eclipse obtained from the fit to the data ${\hspace{3.4cm}}~$
  \end{tabular}
  \end{center}

\end{table*}

The presence of an X-ray pulsar in LMC\,X-4, spinning at $P\simeq13.5$\,s was
revealed with the {\it SAS-3} observatory during a series of X-ray flares
occurred between 1975-1976 \citep{kelley83}. This led to the association of
LMC\,X-4 with the class of the so-called accretion powered X-ray pulsars
(APXPs). Many sources in this class are known to display spin-up episodes,
during which the spin period of the neutron star {decreases}, and
episodes of spin-down, when the rotation of the neutron star is slowed down
probably due to the accretion torques and the interactions between the compact
object magnetosphere and the surrounding accretion disk. Depending on the
different systems, the spin-up and spin-down episodes can last from
days to weeks, extending up to several years in the most extreme cases \citep[see, e.g., reviews of][]
{nagase89,lut94,bildsten97,lutovinov09}.

In this paper, we use all historical spin period measurements of the
neutron star hosted in LMC\,X-4 in order to investigate for the first time the
long-term trends in its variations over the past $\sim$43~yrs. A description
of all the historical measurements is provided in Sect.~\ref{sec:data},
together with the summary of all new determinations of the source spin period
obtained with the most recently available data. We have made use in this
paper of data from twelve different observatories and most of their on-board
instruments. In several cases, the measurements of the source spin period
were taken directly from the literature, while in many others we reanalyzed
the data to improve previous results. A summary of all our findings is
reported in Sect.~\ref{sec:results}. We discuss all the results in
Sect.~\ref{sec:discussion}.

\section{Data sets on LMC\,X-4}
\label{sec:data}

All measurements of the spin period of the neutron star hosted in LMC\,X-4
available to date and used in this paper are summarized in
Table\,\ref{tab:history_tbl}, together with the corresponding uncertainties.
We illustrated in all the following sub-sections how the different
measurements have been obtained. Note that the uncertainties on the spin
period values obtained by us and not retrieved in the literature were
estimated by using the technique described in \citet{boldin13}. For each
observation, we generated a set of $10^4$ synthetic lightcurves in which the
count-rate of the source was randomly varied in each time bin within the
1~$\sigma$ confidence level of the original measurement. We used the mean
value of the pulse period distribution obtained from the synthetic
lightcurves as the best value of the source spin period and the standard
deviation of the same distribution as the associated 1~$\sigma$ uncertainty.

\subsection{{\it SAS-3}}
\label{sec:sas3}

The third US Small Astronomy Satellite ({\it SAS-3}) observed LMC\,X-4 in two
occasions: during 6~days in February 1976 with the Rotating Modulation
Collimators \citep[RMC operating in the 2-11~keV energy band;][] {doxsey76}
and during 1.5~days in May 1977 with the Y-axis detectors \citep[operating in
the 6-12~keV energy band;][]{lewin76}. A total of five flaring episodes
(three in 1976 and two in 1977) were detected during these observations.
X-ray pulsations were registered only during these events. All details of the
data analysis and pulse period measurements are given in \citet{kelley83}. We
quote in Table~\ref{tab:history_tbl} the relevant values from the above
paper.

\subsection{{\it HEAO-2}}
\label{sec:heao2}

LMC\,X-4 was in the field of view of {\it HEAO-2}, the second High Energy
Astrophysical Observatory \citep[{\it Einstein},][]{giacconi79}, several times
during the mission lifetime (from November 1978 to April 1981). The analysis
of these data and the corresponding results were reported in
\citet{naranan85}. Pulsations from the source were significantly detected
only during the X-ray flares occurred on December 16, 1980 (MJD\,44589.6). In
that occasion, the measured apparent spin period was $13.530\pm0.004$ s (not
corrected for the orbital motion). In order to derive the corrected period
we used the orbital ephemerides published by \citet{kelley83} and obtained
the intrinsic period of the neutron star rotation for the date of
the {\it HEAO-2} observations as $P=13.5113$~s. This value is reported in
Table~\ref{tab:history_tbl}. We note that this value does not change
significantly if the more updated ephemerides provided by \citet{molkov15}
are used for the calculation. In this case we obtain $P=13.5117$~s.

\subsection{{\it EXOSAT}}
\label{sec:exosat}

Pulsations from LMC\,X-4 were detected by the European X-ray Observatory
Satellite ({\it EXOSAT}, \citealt{andrews84}) {in five observations
carried out in October-November 1983}. Note that pulsations were
detected not only during the flaring activity of the source but also in its
persistent state during {the more intense observational monitoring of the source
carried out from 1983  November 17 to November 19}. The results of this analysis were
reported by \citet{pietsch85} and included in Table~\ref{tab:history_tbl}.

\subsection{{\it Ginga}}
\label{sec:ginga}

The {\it Astro-C} observatory ({\it Ginga}) was the third Japanese X-ray
astronomy mission \citep{makino87}. The pulsated signal from LMC\,X-4 was
clearly detected during two long observations performed {with the Large
Area Counter instrument \citep[LAC,][]{turner89} on 1988 March 7-10
\citep[][]{woo96} and 1989 August 3-5 \citep[][]{levine91}, when the source
was in a high state of its super-orbital cycle. The pulse period values
reported in these papers are quoted in Table~\ref{tab:history_tbl}}. Note
that the data collected during the source X-ray flares occurred in the
two observations were excluded from the analysis.

\subsection{{\it ROSAT}}
\label{sec:rosat}

Observations of LMC\,X-4 were performed with the Position Sensitive
Proportional Counter \citep[PSPC,][]{pfeffermann87} on-board the Roentgen
Satellite {\it ROSAT} \citep[][]{trumper83} starting from 1991 October 28 to
November 3. Details of the extraction and timing analysis of the source
lightcurves in the $0.1-2.5$~keV energy range are given in \citet{woo96}. In
Table~\ref{tab:history_tbl} we quote the source pulse period value reported
in this paper. Also in this case, the data used to determine the spin period
did not include any flaring episode.

\subsection{{\it ASCA}}
\label{sec:asca}

The Advanced Satellite for Cosmology and Astrophysics ({\it ASCA},
\citealt{tanaka94}). observed LMC\,X-4 on three occasions: 1994 April 26-27,
1995 November 24-25, and 1996 May 24-25. During all observations the source
was in its high super-orbital state and no flares were detected. For the
timing analysis only the first and the third observations were used, as
during the second one the source was caught in eclipse. A first estimate
of the pulse period obtained from these two observations was reported by
\citet{vrtilek97}. These authors showed that the source pulse period was
$13.5069 \pm 0.0007$ in 1994 and $13.5090 \pm 0.0002$ in 1996. These values
were later improved by \citet{paul02a} and we {thus used these updated results}
in Table~\ref{tab:history_tbl}.

\subsection{{\it BeppoSAX}}
\label{sec:sax}

LMC\,X-4 was observed twice with the Italian-Dutch mission {\it BeppoSAX}
\citep[see, e.g.,][and references therein]{piro95} during the 6 years of its
scientific operations. Only in one of two observations (on 1988 October 20-22)
the source was caught in the high super-orbital state and X-ray
pulsations {were recorded}. All details of the {\it BeppoSAX} data
analysis and the measured value of the pulse period were reported by
\citet{naik04}. No X-ray flares were observed in the {\it BeppoSAX} observations of
LMC\,X-4.

\begin{table}
  \centering
 \caption{Log of all {\it RXTE} observations used in this work.}
  \label{tab:rxte_data_tbl}
  \hspace{0.cm}\begin{tabular}{lccc} 
    \hline
    \hline
    Series  & Dates & Total & Num. of\\
    $$ & MJD  & exposure (ks) & segments\\
    \hline
    1 & $50314.6-50316.1$ &  $\sim 60$ & 17 \\
    2 & $50740.3-50749.0$ &  $\sim 31$ & 12 \\
    3 & $51106.0-51114.5$ &  $\sim 128$ & 46 \\
    4 & $51531.5-51533.2$ &  $\sim 98$ & 20 \\
    \hline
  \end{tabular}
\end{table}

\subsection{{\it RXTE}}
\label{sec:rxte}

LMC\,X-4 was observed with the instruments on-board the Rossi X-ray Timing
Explorer \citep[{\it RXTE},][]{bradt93} several times during the mission
lifetime. In order to measure the source pulse period with a reasonable
accuracy, we only made use of the {\it RXTE} data which closely covered a
number of contiguous binary orbits. These sub-set of the {\it RXTE} data
comprised anyway {several dozens of data segments with a typical
duration of few ks}.

We considered for our timing analysis only data from the Proportional Counter
Array \citep[PCA,][]{jahoda06} collected in different event-modes. We
corrected the on-board arrival time of each X-ray event to the rest frame of the Solar System
and filtered out events with an energy below 9\,keV and above 20\,keV. In the 9-20~keV energy band
the pulse profile of LMC\,X-4 is known to be characterized by a simpler shape
\citep[see e.g.][]{levine00}, thus making it easier and more reliable to carry out
a pulse period search. The data containing eclipses and strong X-ray flares
were excluded from the analysis, resulting in a total of four series
of PCA observations, as indicated in Table~\ref{tab:rxte_data_tbl}. We used
the {\sc efsearch} tool within the {\sc FTOOLS} software package to determine
the apparent source pulse period in each of the PCA observations. The
intrinsic pulse period was then calculated by applying {the appropriate corrections
with the updated binary orbit parameters \citep{levine00}}.

\subsection{{\it XMM-Newton}}
\label{sec:xmm}

We have used also data collected with the EPIC-pn and EPIC-MOS instruments
\citep[][]{struder01, turner01} on-board the {\it XMM-Newton} observatory
\citep[][]{jansen01} during two observations of LMC\,X-4. The first one was
carried out on 2003 September 9-10 for a total exposure time of $\sim 113$~ks. The
second observation was performed on 2004 June 16 and provided a
total exposure time of $\sim 43$~ks. All {\it XMM-Newton} data were reduced and
analyzed by using the Science Analysis System (SAS) version 15.0.0. During
both observations LMC\,X-4 was in a high (bright) state and clearly detected
by the EPIC cameras. Strong X-ray flares were {registered} during the first
observation, and two whole X-ray eclipses were detected in both the first and the second data-set.
In order to optimize the estimate of the source pulse period, we excluded from
our analysis all time intervals corresponding to the flares and eclipses. As
the orbital parameters of LMC\,X-4 could not be independently determined from
the {\it XMM-Newton} data, we used the binary ephemerides reported by \citet{molkov15}
in order to correct the on-board arrival time of each {\it XMM-Newton} X-ray event.
The values of the source spin period we obtained from both
observations with the {\sc efsearch} procedure are reported in
Table~\ref{tab:history_tbl}.

\subsection{{\it Suzaku} }
\label{sec:suzaku}

{\it Suzaku} ({\it Astro-E2}) is the fifth Japanese X-ray astronomy mission
\citep[][]{mitsuda07}. Three pointed observations of LMC\,X-4 were carried
out in 2008 (January 15, February 11, and April 5). Details of the timing
analysis are given in \citet{hung10}. As it is not possible to independently determine the
source orbital parameters by using solely the {\it Suzaku} data, these authors measured the pulse period by
applying the {\sc efsearch} technique to the instrumental lightcurves
corrected for the binary motion using the ephemerides from \citet{levine00}. The
source pulse periods  derived from the three observations are reported in
Table~\ref{tab:history_tbl}.

\begin{figure}
\includegraphics[width=0.99\columnwidth,bb=0 150 575 675,clip]{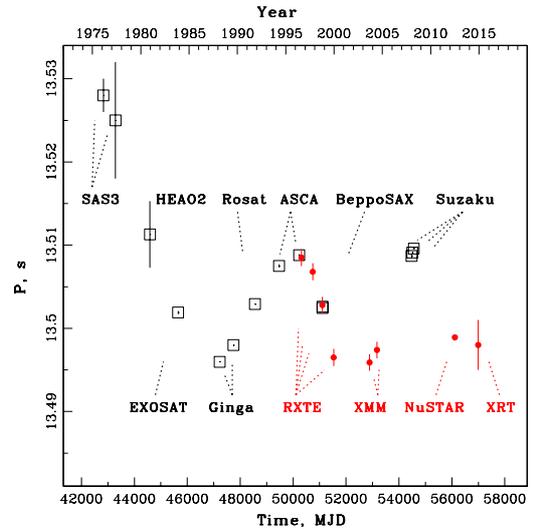}
\caption{Complete history of the pulse periods determined for LMC\,X-4 during
more than 43~yrs of observations. Spin values collected from the literature
are marked with opened-square points. Filled circles are used for the pulse
periods derived in this work. Vertical solid lines represent measurement
uncertainties (in those cases where the lines are not visible the
uncertainties are smaller that the symbol used to mark the measurement).}
\label{fig:history_fig}
\end{figure}

\subsection{{\it NuSTAR}}
\label{sec:nustar}

The Nuclear Spectroscopic Telescope Array ({\it NuSTAR}) mission is the first
focusing high-energy X-ray telescope in orbit \citep[][]{harrison13}. {\it
NuSTAR} is endowed with two co-aligned hard X-ray telescopes (the FPMA and the FPMB)
that operate in the energy band $3-79$ keV. So far, {\it NuSTAR} performed an
observation of LMC\,X-4 on 2012 July 4. The total available exposure time is
$\sim 40$~ks. During the observation the source was caught in the high
super-orbital state and no flaring activity was detected. The source
lightcurves were obtained from both the FPMA and FPMB modules using a
circular extraction region with the radius of $1\arcmin$ centered on the best
known position of LMC\,X-4. The arrival times of the X-ray photons in both lightcurves
were corrected for the Solar System Barycenter and the binary system motion
using the ephemerides from \citet{molkov15}. The source intrinsic pulse
period was determined from these data as $13.49892(3)$~s. {A more detailed
description of the {\it NuSTAR} data analysis, as well as the results of the spectral analysis,
can be found in \citet{shtykovsky16}.}

\subsection{{\it Swift}/XRT}
\label{sec:xrt}

The medium-size mission {\it Swift} is conducting scientific observations in
the X-ray domain since 2004 \citep[][]{gehrels04}. In our analysis we used
data collected by the X-ray Telescope \citep[XRT, see][]{burrows05},
operating in the $0.2-10$ keV energy band. {Our group requested several observations
of LMC\,X-4 during the week spanning from 2014 December 3 to 9}.
The observational campaign consists of a few tens
of pointing, reaching a total effective exposure time of about 30~ks. The source
was observed during one of its super-orbital high state and significantly
detected in each pointing (excluding two of them that caught the source
during the X-ray eclipse). One flaring episode lasting about 15 minutes was
detected during the XRT campaign. {All XRT data used in this paper were carried out
in Windowed Timing (WT) mode}. We excluded from the analysis the time
intervals corresponding to the eclipse and the flaring period. The time
arrival of all photons collected by XRT were corrected to the Solar System
Barycenter before performing any further analysis. The statistics of the XRT
data turned out to be too low to determine the source spin frequency during
the few ks of exposure available for each separated pointing. Therefore we
applied the pulse search procedure ({\sc efsearch}) to the source lightcurve
extracted from the entire observational campaign and corrected for the
binary motion. {The source spin period determined for this campaign is
reported in Table~\ref{tab:history_tbl}.}

\section{Results}
\label{sec:results}

\begin{figure}
\includegraphics[width=0.99\columnwidth,bb=0 150 575 675,clip]{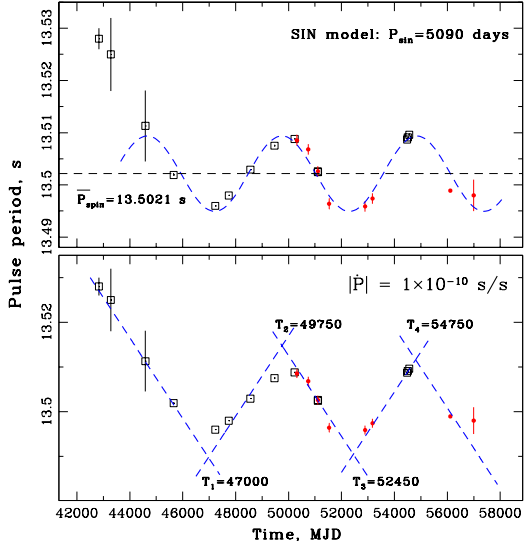}
\caption{Tentative description of the pulse period variations in LMC\,X-4
with two mathematical functions: {the sinusoidal one described by
Eq.~\ref{eq:model_sin} (top panel) and the linear one described by
Eq.~\ref{eq:model_lin} (the bottom panel). See text for details.}}
\label{fig:spin_models}
\end{figure}

The pulse period history of LMC\,X-4 based on all available data mentioned in
the different sub-sections above (from Sect.~\ref{sec:sas3} to \ref{sec:xrt})
is presented in Fig.~\ref{fig:history_fig}. This evolution shows an
intriguing {cyclical} behaviour with torque reversals occurring roughly
every $P_{3}\sim2500$ days (i.e. $\sim$6.8~years). As there is not yet a
clear understanding on the mechanism producing torque reversals in HMXBs, we
did not attempt to describe the data in Fig.~\ref{fig:history_fig} with a physical
model, but rather tried to find some simple mathematical functions that could
qualitatively match the observational results. {Note that the goodness of the
match is not formally evaluated through a proper fit, as the available data are
too sparse. We limited the present attempt to the identification of key
parameters (e.g., the quasi-periodic time-scale of the torque reversals) that can provide some
insights into the mechanisms driving the peculiar variability of the timing properties of the source.}

In the top panel of Fig.~\ref{fig:spin_models} we show the case in which the
apparent periodical variability of the spin period of LMC\,X-4 is tentatively
described with the sinusoidal function:
\begin{equation}
    P(t)=\overline{P}_{spin}\left(1+A{\rm sin}\left( 2 \pi \frac{t-T_0}{P_3}\right)\right).
\label{eq:model_sin}
\end{equation}
Here $\overline{P}_{spin}$ is the mean value of the source spin period over
the whole observational period (in seconds), $P_{3}$ is the characteristic
timescale of the variability, $T_0$ is the time of phase zero, and $A$ the
amplitude of the variability ($t$, $T_0$, and $P_3$ are all in units of
days).

The equation above describes the data reasonably well, and we obtained
$P_3\simeq5090$, $A=5.3\times10^{-4}$\,s, and $\overline{P}=13.5021$\,s. Only
the first two points obtained from the {\it SAS} observatory deviate from the
qualitative periodical function. It is interesting to note that only these
measurements together with the one based on the {\it HEAO-2} data, were
obtained {during episodes of the source flaring activity}. All other spin
period values were measured during quieter periods in the high super-orbital
phase. It is thus possible that some mechanism is at work during the flares
to produce changes in the observed source pulsation period.

As the available observational data (Fig.~\ref{fig:history_fig}) are rather
sparse, we cannot exclude that the different spin-up and spin-down episodes
alternate in a more complicated fashion than a simple sinusoidal trend.
Therefore we show in the bottom panel of Fig.~\ref{fig:spin_models} an
another example in which data are described by a sequence of power laws
with identical absolute values of the slope but alternated signs:

\begin{equation}
    P(t)=P_i-(-1)^i \, |\dot P| \, (t-T_i),  i \in [1,4].
\label{eq:model_lin}
\end{equation}
In the equation above, $\dot P$ is the period derivative, $T_i$ is the time
interval preceding $\it t$ when the pulse-period derivative changes sign, and
$P_i$ is the pulse period in the time interval $T_i$ (in the equation all
time intervals are expressed in days and all spin period values in seconds).
According to this model, the characteristic rate of the pulse-period change
would be $|\dot P|\sim 10^{-10}$\,s/s.

\section{Discussion}
\label{sec:discussion}

The results presented in the previous section show that LMC\,X-4 undergoes
{cyclical} transitions between spin-up and spin-down phases, with
torque reversals occurring roughly every $\sim$6.8~years. The spin period
variability could be reasonably well described by using either a sinusoidal
function, which would suggest the presence of smoothly periodical variations
between two values of maximum and minimum spin periods, or a slightly more
complicated function, relaxing the needs of assuming a strictly periodical
phenomenon. At present, given the irregular and large spacing among all data
available, it is not possible to firmly distinguish among these two
possibilities. Nevertheless, below we discuss briefly several possible
mechanisms that could give rise to quasi-periodical torque reversals in
systems like LMC\,X-4.

{\it Third body.} Concerning the strictly periodical case (i.e.
Eq.~\ref{eq:model_sin}), one of the most intriguing possibility is that the
sinusoidal-like behaviour of the pulse period are due to the Doppler shifts
produced when a binary system is orbiting around a third body. In this case,
we can re-write Eq.~\ref{eq:model_sin} as
\begin{equation}
    P(t)=\overline{P}_{spin}\left(1+\frac{K_{B}}{c}{\rm sin}\left( 2 \pi \frac{t-T_0}{P_3}\right)\right),
\label{eq:model_sin2}
\end{equation}
\noindent
where $K_{B}$ is the amplitude of the radial velocity of the binary system
and $c$ is the speed of light. By using the results obtained in
Sect.~\ref{sec:results}, we can estimate an orbital period of the inner HMXB
around the third body of $P_3\simeq5090$~days and a radial velocity of the
binary of $K_B=160$~km~s$^{-1}$. The mass of the third body, $M_3$, can be
estimated from these parameters by using the ``mass function'':
\begin{equation}
    M_3=\frac{K_B^3 P_3}{2 \pi\,G\,{\rm sin}^3(i)}(1+q)^2
\label{eq:mass_det}
\end{equation}
where $q=M_B/M_3$, $M_B$ is the total mass of the binary system, $G$ is the
gravitational constant, and $i$ is the inclination of the system (we assumed
here the case of circular orbits for simplicity). Assuming $i=90^o$ and $q
\ll 1$, we can estimate from Eq.~\ref{eq:mass_det} a lower-limit for the mass
of the third body of $M_3\simeq2000~M_\odot$.

Although the above finding would open the interesting possibility of LMC\,X-4
being part of a system hosting an intermediate mass black holes (IMBHs), we
show in the following that this interpretation can be most likely ruled out.

\begin{figure}
\includegraphics[width=0.99\columnwidth,bb=0 150 575 675,clip]{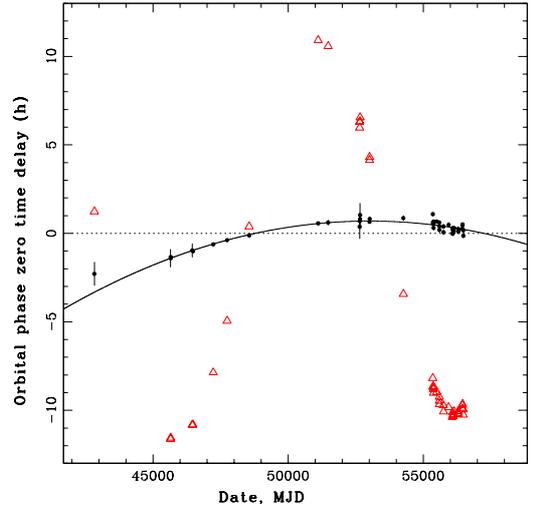}
\caption{Light travel time delays obtained
from all available mid-eclipse epochs measured throughout the historical
X-ray observations of LMC\,X-4 \citep[all points represented with black dots
are obtained from][]{molkov15}. The solid curve represents the best fit to
the data with a quadratic model that takes into account the orbital evolution
of the system \citep{molkov15}. Red triangles show the light travel time
delays expected in case the binary system was orbiting around a third body
with a period $P_3\simeq5090$~days.} \label{fig:pca_third}
\end{figure}

The binary system LMC\,X-4 is located at a distance of $\simeq4$ kpc from the
dynamical center of the Large Magellanic Cloud. The structure of the galaxy
and its kinematics are well studied and according to relatively recent works
\citep[see, e.g,][]{marel06, marel14} the line-of-sight velocity of stars
close to LMC\,X-4 with respect to the Solar System is about 300 km\,s$^{-1}$.
We found in the literature three direct measurements of the mean
line-of-sight velocity $V_{LOS}$ of LMC\,X-4 in different epochs: (i)
$V_{LOS}=294\pm5$ km\,s$^{-1}$ on MJD\,$43198$ \citep[][]{chevalier77}, (ii)
$V_{LOS}=302\pm13$ km\,s$^{-1}$ on MJD\,$43475$ \citep[][]{hutchings78}, and
(iii) $V_{LOS}=306\pm10$ km\,s$^{-1}$ on MJD\,$52232$ \citep[][]{meer07}.
These measurements were obtained from the analysis of the Doppler shift of
absorption and emission lines registered in the optical spectrum of the donor
star in this system. All these values are in a good agreement with the radial
velocity distribution in the LMC galaxy and cannot easily be reconciled with
the expectations of the triple system model. In the latter case, Doppler
effects are expected to produce shifts of $V_{LOS}$ at different orbital
phases. The measurements (i) and (ii) mentioned above would correspond to the
two phases of the triple system orbit $\Psi_3^{II}\simeq0.52$ and
$\Psi_3^{III}\simeq0.23$. At these phases, the expected Doppler shifts of
$V_{LOS}$ would be $\Delta V_D^{II}=+20$ and $\Delta V_D^{III}=-159$
km\,s$^{-1}$, respectively. This should lead to a discrepancy between the two
measurements at MJD\,$43198$ and MJD\,$43475$ of about $\sim
180$~km\,s$^{-1}$, which is not observed (see also
Fig.~\ref{fig:pca_third}).

As an additional test of the triple system hypothesis, we considered all the
mid-eclipse times measured by different instruments throughout the entire
history of X-ray observations of LMC\,X-4 \citep[][]{molkov15}. If the source
was orbiting a third body, sinusoidal-like residuals would be expected after
the long-term evolution of the orbital period is removed from the fit to the
data. The plot in Fig.\,\ref{fig:pca_third} shows that this is not the case.
In the same figure we also show that the amplitude of the sinusoidal
modulation expected for a third body with the orbital period $P_3$ estimated
in Sect.~\ref{sec:results} would be much larger than the uncertainty on all
available measurements. Indeed, based on our previous calculations in this
section, the orbital radius of the third body should be $R_3=K_B\,P_3/(2
\pi)\simeq 1.2\times 10^{10}$\,km, leading to a maximum light travel time
delay of $\sim 10.3~h$.

{\it Variations of the accretion rate.} The hypothesis discussed above implies
apparent changes in the spin frequency. At the same time in binary systems there are
mechanisms and processes that could directly change the rotation rate of a neutron star.
The most important of them is connected with the transfer of the angular momentum between
the accreted matter and the neutron star, i.e. the ``accretion torque''. The balance between these
concurrent processes depends on many factors, the dominant one being usually the accretion rate.
It is generally assumed that the neutron star in LMC\,X-4 accretes from
a disk, and thus a number of different models for the estimate of the accretion torque
can be adopted to estimate the expected spin-up/spin-down rate
\citep[see, e.g., ][, etc.]{1979ApJ...234..296G,lovelace95,1995ApJ...449L.153W}.
In the majority of these models, a substantial change in the spin of the neutron star
requires a large variation of the mass accretion rate, which in turns leads to
comparable changes in the X-ray luminosity.
This is at odds with respect to the findings obtained from the long-term observations of
LMC\,X-4, which show that the source does not display a significant variation in the
X-ray luminosity during the torque reversal events.

{\it Recycling magnetosphere model.} A different possibility to interpret the
regular torque reversals observed from LMC\,X-4 without assuming a strong
variations of the mass accretion rate is offered by the so-called ``recycling
magnetosphere model'' presented in \citet{perna06}. These authors showed that
when the magnetic axis of the compact object is inclined with respect to its
rotational axis and the direction normal to the plane of the disk, accretion
onto the neutron star cannot proceed smoothly. As for an inclined dipole the
magnetospheric boundary regulating the interaction between the compact object
and the accretion disk is elongated, it might occur that accretion only takes
place during a few spin rotational phases, while in other the inflowing
material is preferentially ejected from the system due to the onset of the
propeller effect. Depending on the strength of the propeller, part of the
ejected material can be unbound completely from the system or fall back onto
the disk and contribute to enhance the accretion at a later stage onto the
neutron star. Different model parameters (the neutron star spin period,
magnetic field strength, inclination angle, and mass accretion rate) can give
rise either to a system in which accretion prevails for most of the time and
the neutron star shows a strong spin-up phase, or systems in which the
ejection/recycling are the predominant interaction mechanisms and a torque
reversal to a spin down phase is observed.

\citet{perna06} showed that, even with a constant mass inflow rate, the
secular effects produced by a recycling magnetosphere would induce periodical
switches between spin-up and spin-down episodes in a quasi-periodical
fashion. The frequency of the torque reversals depend on all system
parameters, but range typically from few to tens of years due to the large
moment of inertia of the neutron star. In a few cases, torque reversals have
been predicted to occur at constant X-ray luminosities (see, e.g., the case
of 4U\,1626-67), while in others a significant change in the overall X-ray
intensity of the source was observed (e.g., GX\,1+4). This matches the
observational results reported in the literature for a number of X-ray
pulsars in binary systems \citep[see, e.g.,][and references
therein]{chakrabarty97,fritz06,camero-arranz10}. The addition of a long-term
modulation of the mass accretion rate could also produce additional
complications in the resulting sequence of torque reversals, as well as in
the X-ray luminosity variability occurring at the time of the
spin-up/spin-down transitions. A more quantitative comparison between the
torque reversals observed from LMC\,X-4 and the predictions of the recycling
magnetosphere model requires the analysis of the source luminosity across all
historical X-ray observations and particularly during the torque reversal
episodes. As these measurements are strongly dependent from the responses of
the different instruments, a different re-analysis of all data sets is
necessary to obtain fully comparable values. This is beyond the scope of the
present work.

{\it Different states of the neutron star magnetosphere.} Another possibility
to interpret the quasi-periodical variations of the pulse period observed in
LMC\,X-4 is related with processes in the interior of the neutron star
or within its magnetosphere. Slow variations of the spin-down rate on time scales ranging
from months to years, i.e. the so-called ''timing noise'', have been
observed from isolated neutron stars practically since
their discovery (mainly in the radio domain). The shape of these low-frequency variations
changes with time and resemble a quasi-periodical signal \citep[for review see, e.g., ][and
references therein]{hobbs10} or even a more strictly periodical modulation
\citep[][]{kerr16}. This is closely reminiscent of what we are observing from LMC\,X-4.
The physical phenomenon causing the timing noise has not been well established yes.
One of the most credited possibility is that of the so-called ''state switching'' model,
where the timing noise is driven by changes in the pulsars magnetosphere
\citep[][]{lyne10}. This model assumes that there are two (or more) states of
the neutron star magnetosphere with different parameters regulating the
spin-down rate. The neutron star can persist in any of these states even for a
long time but the switch between states occurs abruptly. Assuming that such processes
take place also in LMC\,X-4, it is reasonable to assume that changes in the coupling
between the neutron star and the disk in the different states can lead to
substantial changes in the pulse period. Even though this is only a speculative idea
and more refined investigations are needed to eventually confirm its validity, we note that
recent observations of the transient X-ray pulsar V\,0332+53 suggested that
possible switches between different interaction states of the neutron star magnetosphere and the
accretion disks can take place in X-ray pulsars \citep{doroshenko16}.

\section*{Acknowledgements}

This work was supported by the Russian Science Foundation
(grant 14-22-00271). We thank D.Yakovlev and K.Postnov for
useful discussions of the obtained results.













\bsp	
\label{lastpage}
\end{document}